# Influence of nanoparticles mechanical properties on macrocracks formation

*Ludovic Pauchard[1], Bérengère Abou[2], Ken Sekimoto[1,2],*

[1] Laboratoire Fluides, Automatique et Systèmes Thermiques, Université Paris VI, Paris XI, UMR CNRS 7608, France

[2] Laboratoire Matière et Systèmes Complexes, UMR 7057 CNRS & Université Paris Diderot, France

[3] Gulliver, UMR7083 CNRS & ESPCI, France

**We present an experimental investigation of drying suspensions of both hard and soft nanolatex spheres. The crack formation is examined as a function of the proportion of hard and soft deformable particles, leading to tunable elastic properties of the drying film. In our experiments corresponding to a given layer thickness, no crack formation could be observed below an onset value of the proportion in hard spheres $\phi \sim 0.45$. During the drying process, the mass of films with various compositions in hard and soft spheres is measured as a function of time. The results suggest that the soft particles undergo deformation that releases the internal stresses. The spacing between cracks is measured as a function of the layer thickness for various compositions in hard and soft spheres.**

**Introduction**

The use of coatings in most applications is based on the formation of a continuous film during drying. The industrial applications of these films depend on their final mechanical properties but the film formation is a complex process and is considered as a succession of several steps such as concentration of dispersion, particle packing, particles deformation,... During consolidation, cracking and warping may occur that generally need to be avoided. The phenomenon of cracking in drying films depends on several parameters[1,2]. Some of them control the boundary conditions of the film : substrate controlling the adhesion of the film ; ambient temperature and humidity rate controlling drying kinetics a the surface of evaporation. Others are due to the film itself : film thickness, particle size and mechanical properties of the matrix formed by the closed packed array of particles[3,4,5].

In this paper, we suggest a model system to study and quantify the crack morphology observed in elastic or visco-elastic systems, such as painting layers. It consists in a drying film of a dispersion of both hard and soft polymer spheres in a controlled proportion. The influence of the matrix mechanical properties on the final cracks patterns formed in the film, are investigated. As a latex film-forming dries, it is transformed from a colloidal dispersion into either a porous matrix or a continuous polymer film, both having mechanical integrity. The drying curves –mass and drying rate- are recorded as a function of time. The nature of the latex film-forming obtained at the end of the drying process depends on the glass transition temperature of the particles and the ambient temperature. If the glass transition temperature of the latex particles is higher than the ambient temperature (hard particles), the final film is found to contain a large number of cracks in it. Contrarily (soft particles), the final film may be homogeneous[6]. We measure the crack spacing as a function of the layer thickness and composition of various suspensions of nanolatex spheres.

It has been believed that, when the film is applied on a glass substrate, evaporation of the solvent concentrates the particles into a close packed array, before the film cracks to release the stresses. Our measurement of the mass and the inter-crack spacing shows, however, that the evaporation from the surface is virtually blocked in the presence of soft particles, and that, even in the case of solely hard spheres, there is more inter-particle space than the close packed state upon the first crack appears.

**Materials and methods**

The experiment consists in a drying thin film of a dispersion of both hard and soft nanolatex spheres in a variable and controlled proportion. The colloidal spheres, with two different glass transition temperatures $T_g$, and made of a styrene copolymer and acrylic acid (provided by Rhodia Recherche, Aubervilliers, France), are *30 nm* in diameter. At room temperature, this allows us to work with hard spheres ($T_g$ = *100 °C*) and soft spheres ($T_g$ = *0 °C*) that can deform more easily than hard spheres. The initial volume fraction of both dispersion is $\phi_0$=*0.3*, and the density of pure dry material of hard and soft spheres is respectively $\rho_h$ = *1.08 ± 0.02 g.cm$^{-3}$* and $\rho_s$ = *1.10 ± 0.02 g.cm$^{-3}$*. The proportion $\phi$ in hard spheres is defined as $\phi = V_h/(V_h + V_s)$, where $V_{h(s)}$ is the volume of hard (respectively soft) spheres. Binary mixtures are magnetically stirred during *15 min* at room temperature, and then sonicated at a *1 s* on/off pulse interval during *300 s*. When there is no evaporation, the suspensions are stable. The experiments are conducted at room temperature and relative humidity around *37 ± 2 %*. In our experiments, the transfer of water in air is limited by diffusion and therefore controlled by the relative humidity RH. The surface tension of each suspension is measured by the Wilhelmy plate method and displays a nearly identical value of *57 ± 5 mN/m*.

The dispersion under study is deposited in a circular container which is *16 mm* in diameter, with Altuglas lateral walls, and a glass substrate (Figure 1). The contact line of the dispersion remains quenched at the upper edge of the container during the drying process. The drying layer thickness is thus constant in the container center, covering about *70%* of the total surface area, allowing for homogeneous evaporation, and in some cases to isotropic crack patterns. Starting from the beginning of the drying process, the mass of the container filled with the suspension, is measured with a precision balance (Sartorius). The mass $m$ of films with different compositions in hard and soft spheres was recorded during the drying process. The drying rate $|dm/dt|$ was then extracted from the mass measurements. Simultaneously, the cracks morphology is recorded with a Leica camera positioned on the top of the sample. At the end of the drying process, a film with a thickness $H$ is obtained. The final film thickness is measured by successively focusing a microscope objective on the substrate and then the surface layer, with an estimated precision of $2\mu$m.

The macroscopic elastic response of the colloidal gel is characterized using the CSM Instruments Micro Indentation testing (MHT) by measuring creep. The method consists in maintaining an applied force at a constant maximum value and measuring the change in depth of the indenter as a function of time. The relative change of the indentation depth is referred to as the "creep" of the specimen material.

The volume fraction of the solvent $\phi_w(t)$ at time $t$, is deduced from the measurement of the sample mass $m(t)$, assuming that (1) only the solvent evaporates from the sample, and (2) no air penetrates into the sample up to time $t$ (see reference[7]), (3) the volumes occupied by hard and soft particles are constant throughout the drying process (even in case of deformation of spheres). At time $t$, the sample mass $m(t)$ simply writes $m(t) = \rho_w V_w(t) + \rho_h V_h + \rho_s V_s$, where $\rho_{w\,(h,s)}$ is the density of water (respectively hard and soft spheres), and $V_{w\,(h,s)}$ the volume occupied by water (respectively hard and soft spheres). Given a mixing ratio $\phi$, the volumes occupied by hard and soft spheres can easily be calculated from the known quantities $m(0)$, $V_{tot}(0)$, $\phi$ and $\phi_0$. The solvent volume fraction $\phi_w(t) = V_w(t)/V_{tot}(t)$ can then be expressed implicitly as :

$$\phi_w(t) = \frac{1}{V_{tot}(t)}\left[\frac{m(t)}{\rho_w} - V_h\frac{\rho_h}{\rho_w} - V_s\frac{\rho_s}{\rho_w}\right] \qquad (1)$$

where $V_{tot}(t) = \phi_w(t)V_{tot}(t) + \rho_h V_h + \rho_s V_s$.

Note that under the assumption (2), and in case the particles are not deformed from the spherical shape, the value $\phi_w(t)$ should be smaller than $1 - \phi_{rcp}$, where $\phi_{rcp}$ ($\approx 0.64$) is the random close packing volume fraction.

**Results**

Figure 2 shows the drying curves, mass $m$ and drying rate $|dm/dt|$, of suspensions with different compositions, as a function of a dimensionless time $t/t_D$, where $t_D = H_0/v_e$ denotes the characteristic drying time, $H_0$ the initial layer thickness and $v_e$ the evaporation rate of water in the air (typically $v_e = 10^{-4}$ mm/s at $RH = 40\%$). This dimensionless time allows us to balance for the slight differences in relative humidity between the curves.

In the particular case of a suspension of hard spheres ($\phi=1$), the drying rate change with time has been separated into three periods, to which physical processes have been usually associated[1,8,9]. During the first regime associated with the existence of a continuous liquid network through the sample, the solvent evaporates at a constant rate from the free surface (region I in the insert in Figure 2). Then the particles are dynamically arrested. The resulting porous matrix, saturated with solvent, is under compression at its free surface, due to high capillary pressure. However, this shrinkage is limited by the adhesion on the solid substrate which limits the contraction of the gel. This frustration causes mechanical tensions that build up with time. The constant rate drying period is followed by a non-linear drying rate period where the drying rate decreases (regime II). The free water still evaporates from the surface where the liquid is carried from the inside by the fluid flow. During this stage, the particles mechanical properties are of importance since the particles may deform elastically under high capillary pressure, and thus close the inter-particles voids. During the final regime (III), at some point the liquid phase breaks up into separate fractions and drying is said to enter the second "falling rate period" completing the drying process. The associated drying rate is much slower than those of the two first regimes.

In the case of mixtures of hard and soft spheres, these three regimes can be identified as well, as can be seen in Figure 2 ($\phi \neq 1$). Besides, the drying process is found also to strongly depend on the film composition in both hard and soft spheres. During the first stage, the evaporation rate remains independent of the film composition. However, slower evaporation of the solvent could be measured during regime II, in the presence of soft spheres ($\phi \neq 1$). The more the film contains soft spheres, the slower the evaporation is, as can be clearly seen in Figure 2. At the end of the measurements, the final mass of the films containing soft spheres is larger than in the case of hard spheres films. Because hard and soft spheres have same diameter and density, these results suggest that water remains trapped in the films containing soft spheres. This strongly suggests that during the film drying, soft particles can deform, causing the inter-particle voids to close and block further evaporation. The more soft particles in the suspension, the more inter-particle voids can be closed by soft particles deformation, resulting in earlier stopping of evaporation with increasing soft sphere proportion[10].

Moreover, for $\phi$ high enough ($\phi > 0.45$ for a final thickness $H = 170$ $\mu m$), the stress built up results in cracks formation. The experiments show that cracks start invading the film in the beginning of regime II (insert in Figure 2). Measurements of the time-lag cracking, $t_{crack}$, at which cracking occurs for gel of various compositions $\phi$, are shown in Table 1. As can be seen, the more the film contains soft spheres the more delayed cracking occurs. An estimate of the water volume fraction upon the first appearance of cracks, $\phi_{w,crack} \equiv \phi_w(t_{crack})$, can be deduced from expression (1) taking into account the total mass of the sample at $t = t_{crack}$ measured from the mass variations with time (Figure 2). The solid volume fraction 1-$\phi_{w,crack}$ can then easily be deduced from the data. The value of $\phi_{w,crack}$ decreases with decreasing proportion of rigid spheres $\phi$ (Table 1). The maximum value $(\phi_{w,crack})_{max} = 0.40 \pm 0.03$ (when cracking occurs), is obtained when the film is purely composed of rigid spheres, $\phi=1$. Experiments show that cracking occurs earlier in time in rigid films, and in this case, the volume fraction of water reaches a maximum value. In the case of a relatively soft film ($\phi$ = 0.52), $\phi_{w,crack}$ = *0.30 ± 0.03* is smaller than 1− $\phi_{rcp}$ = 0.36, which suggests that some of the inter-particle voids in the film are closed if the assumption (2) mentioned above remains valid. This indicates therefore that some of the soft particles are deformed from their spherical shape.

The aspects of drying films with different composition in both hard and soft spheres are investigated. Figure 3 presents the results obtained for films with a final thickness *H = 170 ± 5 µm*. At the end of the experiments, the mean cell area is measured as a function of the film composition $\phi$ in hard and soft spheres. Below a threshold value in hard spheres *$\phi$ ~ 0.45*, no crack formation could be observed and the film remains homogeneous in the horizontal directions. Around *$\phi$ ~ 0.45*, cracks nucleate in the sample. The cracks growth stops shortly after their initiation leading to isolated crack segments. They are either open ended single cracks or star like patterns with three cracks at *120°* from each other and centered at defects as nucleation sites[11]. At higher proportion in hard spheres $\phi$, cracks may propagate in the sample, connecting each other, and dividing the layer into adjacent cells of the surface area denoted by $A_{cell}$; this quantity will be the parameter of interest in the following, because it can be accurately measured from observations[12]. In Figure 3, the characteristic cracks spacing is found to decrease with the proportion in hard spheres $\phi$. This suggests that more crack generation is needed to relax the mechanical stresses in the film with increasing $\phi$ (Figure 3, *$\phi$~ 0.58, 0.70, 1*).

Moreover, in the range of $\phi$ corresponding to cracks formation, residual stresses in the polygonal cells can induce another mode of crack formation. When the stored elastic strain energy in the film overcomes the adhesion energy on the substrate film, the detachment due to crack formation at the interface takes place between the film and the substrate. Consequently, this process releases in-plane stresses in both in-plane directions. The buckle propagation stops when the elastic strain energy is just overcome by the adhesion energy. As a result, a single adhering region is located in each polygonal cell at the final stage of the process. In the case of film of pure hard particles, the adhering region is preferentially circular at the final stage of the process[13] (region bordered by dashed line inside the polygonal cell in Figure 4). Changing the composition of the film, different morphologies can be observed. At the final stage of the detachment process, the pattern of the adhering region in comparison with the polygonal cell surface can be characterized by the relative variation of the respective surface areas, $A_{cell}$ and $A_{adh}$, also the relative variation of the circularity, $C$, describing the relative change between the perimeter of the adhering area and the polygonal cell area (Figure 4) (note that a circularity value of $1$ indicates a perfect circle; A value approaching $0$ indicates an increasingly elongated polygon). The ratio between the adhering surface area and the polygonal surface area decreases when increasing the composition in soft particles in the film. The buckle propagation is more difficult. Moreover, for a composition of hard particles approaching 0.6, detachment of the film is simply located close to the crack opening (Figure 4).

**Discussion**

Let us first recall a widely accepted macroscopic picture of the mechanism of drying fracture[14]. A colloidal gel layer of infinite horizontal dimensions and of the initial thickness given by the dot-dashed horizontal line in Figure 5a is dried down to the thickness shown by a filled thick rectangle. In the figure the horizontal dimension of the gel is truncated, while the solid substrate (thin rectangle) is somehow prolonged. The material cannot change its dimensions along its horizontal directions because the gel is attached to the substrate. If it were detached, the relaxed state of the dried gel material would have the form represented by the dashed open rectangle in Figure 5a or the two dashed open rectangles in Figure 5b *if* we cut the material into two blocks. The gel attached to the substrate in Figure 5a is, therefore, under horizontal stretching as a whole (to compensate, the substrate is under lateral compression).

In Figure 5b, a crack opens by creating the V-shaped new surface. The tensile deformation of the gel material, within a distance comparable to the thickness, is partially relaxed. This reduces the stored elastic energy. The formation of a crack is favored if this gain of elastic energy overcompensates the cost of creating the crack surfaces. Thus the crack appears one after one. If, however, a new crack were to appear at too small distance from pre-existing ones, the released elastic energy per the last crack would be not enough to compensate the cost of the surface energy of the crack. In this scenario, the final spacing between the cracks is determined accordingly to Russel *et al.*[15].

Recently, there are different models that predict how the elastic properties of the particles influence the creation of cracks upon drying. For colloidal dispersions of hard particles, Man *et al.*[16] showed that, given a thickness of drying film, the gel with harder particles has less tendency to create cracks. Physical reason is that the network of hard particles can store little elastic energy before the capillary (under-)pressure reaches its maximum value. Beyond this limit, the air penetrates into individual inter-particle space without creating macroscopic cracks. Singh *et al.*[17] proposed an alternative case of soft particles and predicted that the crack creation is unfavored for the colloidal dispersions of very soft particles. They argued that, upon drying, the stocked elastic energy is bounded from above if the deformed particles fill the entire space of gel. Both scenarios predict that the crack is easier to be created in thicker films, in accordance with experiments. However, the minimum thickness for the crack creation $H_c$ is predicted to increase with shear elastic modulus of particle $G$ for hard particles and to decrease with $G$ for soft particles. We note that the argument on the crack spacing by Russel *et al.*[15] applies to both scenarios.

The crack spacing of mixed colloidal spheres can be predicted along this line (Russel *et al.*[15]) if we suppose that, from macroscopic point of view, our suspension is a homogeneous material with a tunable rigidity. This rigidity can be quantified by the measurement of the macroscopic elastic response, $\bar{G}$, as a function of the fraction of hard particles, $\phi$, (Figure 6 and Appendix). In our experiments, the film of a fixed thickness developed cracks only when the hard particles occupy a minimum volume fraction. Therefore, our system behaves more like the colloidal suspension of soft particles discussed by Singh and Tirumkudulu[18]. However, if we look into more quantitative aspects, the threshold value of the elastic coefficient, $\bar{G} \equiv \bar{G}^*$, for the appearance of the first crack is larger by 6-7 orders of magnitude than what one can deduce from their paper (eq.(2) therein). The theoretical prediction of the crack spacing, $W$ vs $\bar{G}$ with appropriate non-dimensionalization (Tirumkudulu and Russel[18]) can be compared with our measurement of the space between the cracks, $A_{cell}^{1/2}$, vs the fraction of hard particles, $\phi$, (Figure 3) indirectly if we use our simple superposition approximation of $\bar{G}$ vs $\phi$. The theory predicts the divergence of the spacing to thickness ratio as:

$$\frac{W(\phi)}{H} \approx \ln\left(\frac{2\bar{G}^*}{\bar{G}(\phi) - \bar{G}^*}\right) \tag{2}$$

for $\bar{G}$ slightly above $\bar{G}^*$. The predicted $W/H$ (the solid curve for f > 0.45 in Figure 3), however, depends on $\phi$ much stronger than the measured data.

At the moment we have no concise theory to explain these discrepancies. We should, however, note several differences between our experiments and the assumptions for the models mentioned above. First, our measurement exhibits the blocking of evaporation if the sample contains a lot of soft particles. We, therefore, should take into account the non-quasistatic effects, especially the inhomogeneity of the solvent content along the vertical direction[19]. Secondly, the packing of particles may not be close to the random close packing. According to Table I, the fracture appears earliest with purely hard particles, and upon the appearance of the first fracture the purely hard particles occupy the least fraction of volume (1-0.40 =0.60). While this fraction less than $\phi_{rcp}$ is rather surprising in view of the hypothesis of random close packing in the previous papers, it is consistent with our observation that the sample's linear dimension continues to shrink slowly by ~10% after the appearance of cracks. In other words, the *volumetric plasticity* may be also a factor limiting the stored elastic energy.

**Conclusion**

In this work we have investigated drying kinetics and crack patterns caracteristics of gels as a function of their composition. Whereas the first regime of drying kinetics is not influenced by the composition of the layer, evaporation from the layer surface is blocked in the presence of soft particles. This first result suggests a closing of the inter-particles voids during the porous matrix consolidation. In this hypothesis the process is driven by internal modifications of the film, i.e. soft particles deformation, releasing a part of the elastic energy. As a result the number of cracks decreases with the proportion in soft spheres. The poor quantitative agreements with simple extrapolation on existing theoretical models suggests that the non-equilibrium blockade of evaporation as well as the volumetric plastic compaction towards the closed packing play an important role for the colloid containing soft particles.

ACKNOWLEDGMENT. This work was supported by the ANR Programme Jeunes Chercheurs ANR-05-JCJC-0029. We would like to thank L. Limat for useful discussions and I. Jeté-Côté for the help in the experiments. K. Sekimoto acknowledges the discussion with T. Narita and F. Lequeux.

Appendix : Linear superposition approximation of the elastic response of a colloidal gel

Russel *et al*. (2008)[15] proposed a mean-field approach for the elastic response of a dense packing of spherical particles. They assumed the affine deformation of each pair of contacting particles and average over all orientations of contact surface. They showed that, under the uniform compaction normal to the substrate by a factor *(1-ε₀)*, i.e. under the strain of $\epsilon = -\epsilon_0 \hat{z}\hat{z}$ ($\hat{z}$ being the unit vector normal to the substrate), the stress exerted by the particles network, $\sigma_0$ is given by:

$$\sigma_0 = -p_0 \delta - \frac{\bar{G}}{6}[\delta + 3\hat{z}\hat{z}]\varepsilon_0^{3/2},$$

where $\delta$ is the unit matrix, the hydrostatic compression $p_0$ is the capillary pressure in the case considered. The parameter $\bar{G}$ is defined as:

$$\bar{G} \equiv \frac{\phi_p N}{\pi} \frac{G}{2(1-\nu)}$$

where $G$ and $\nu$ are, respectively, the Young modulus and the Poisson ratio of the particles, $N$ (≈6) and $\phi_p$, respectively, the mean number of contact for a given particle and the volume fraction of the particles in the random packing state. As a straightforward generalization of this approach for the mixture of hard and soft particles, we can replace $G/(1-\nu)$ in the above by the average over the values for the different type of contacts, i.e. the hard-hard, hard-soft and soft-soft pairs. For the hard-hard and soft-soft pairs, we take the values of $G/(1-\nu)$ of the hard particles, $G_h/(1-\nu_h)$, and soft particles, $G_s/(1-\nu_s)$, respectively.

For the hard-soft pairs, we must replace the above $G/(1-\nu)$ by the harmonic average (Johnson, 1985)[19], that is:

$$[G/(1-\nu)]^{-1} \rightarrow ([G_h/(1-\nu_h)]^{-1} + [G_s/(1-\nu_s)]^{-1})/2.$$

The macroscopic elastic response $\bar{G}$ is then obtained when replacing $G/(1-\nu)$ in the above $\bar{G}$ by the average over the three types of contact, such as:

$$G/(1-\nu) \rightarrow \phi^2 G_h/(1-\nu_h) + 4\phi(1-\phi)([G_h/(1-\nu_h)]^{-1} + [G_s/(1-\nu_s)]^{-1})^{-1} + (1-\phi)^2 G_s/(1-\nu_s)$$

where $\phi$ is the proportion of hard particles among all particles. From this expression, it can easily be shown that the presence of soft spheres decreases the value of the effective $\bar{G}$. According the argumentation of Russel *et al.* (2008)[15], the smaller effective modulus $\bar{G}$, the smaller stored elastic energy in the film.

Using these modifications, the theoretical macroscopic uniaxial modulus, $\bar{G}$, is plotted as a function of $\phi$ in Figure 6 (dashed curve), where we used the values, $G_h = 10^4$ MPa, $G_s = 30$ MPa and $\nu_h = \nu_s = 1/3$. Figure 6 also shows the macroscopic elastic response $\bar{G}$ measurements for various compositions in rigid and soft spheres $\phi$, performed by micro-indentation tests.

For $\phi$ below about 30%, both the experiment and theory shows that the magnitude of macroscopic modulus is mostly determined by that of soft particles. However, the measurement exhibits a clear onset phenomena at around 40% from which the macroscopic modulus rises sharply. The latter behavior suggests the percolation of rigidity in the mixture of soft and hard particles.

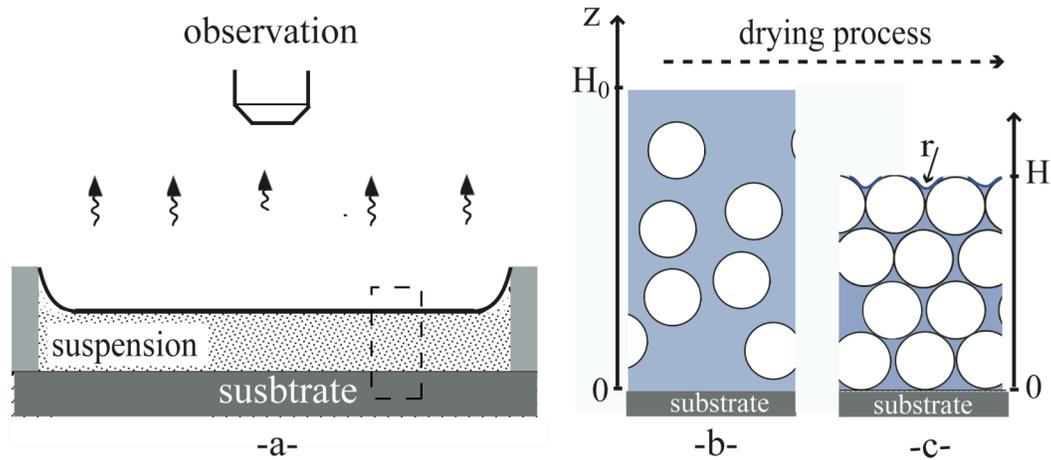

**Figure 1.** (a) Experimental set-up (side view). The suspension is deposited in a circular container which is 16 mm in diameter. Reproducible adhesion of the layer on the substrate is obtained by a carefully cleaned glass slide. Evaporation of the solvent from the surface induces consolidation of the layer, and possibly cracks formation. The top of the sample was recorded during the whole drying process through a Leica microscope. The region shown in the dashed rectangle is sketched in (b,c). Illustration of the film of initial thickness $H_0$ just after the film deposition (b) and when a porous matrix saturated by solvent is formed (c). The curvature $r^{-1}$ of the air-solvent meniscus increases during evaporation.

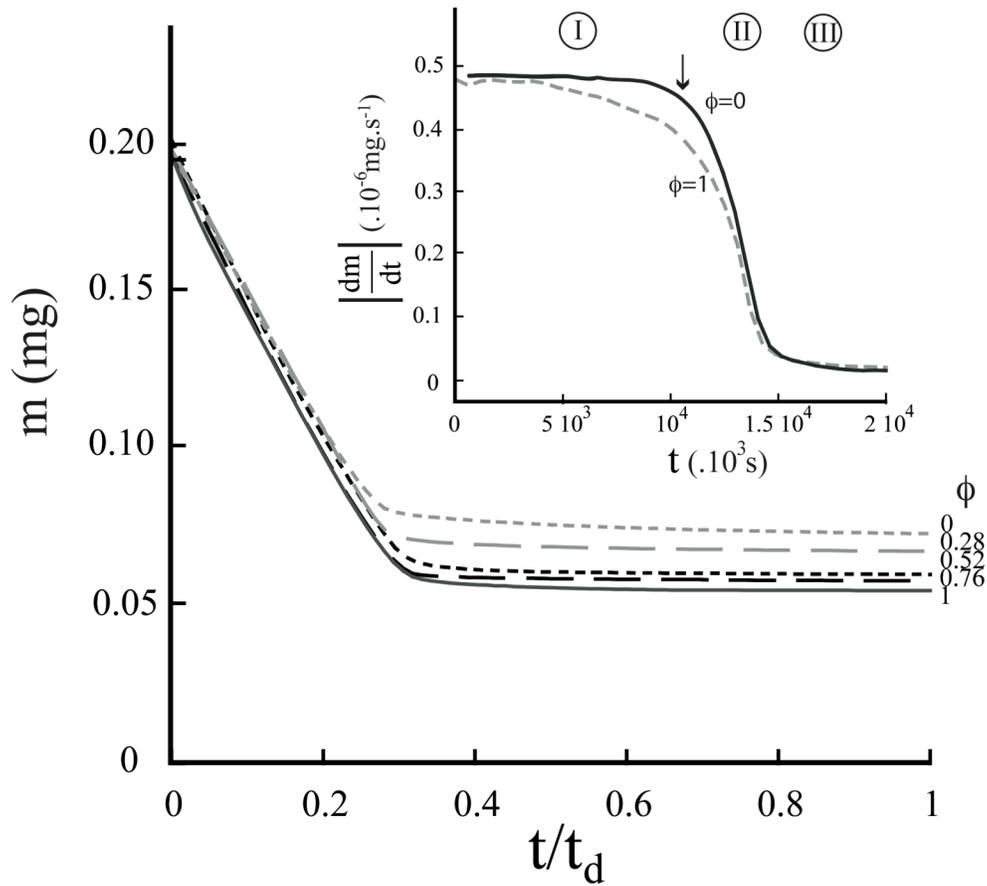

**Figure 2.** Mass of drying films as a function of the dimensionless time, for various compositions in rigid and soft spheres. From top to bottom, the proportion in hard spheres is: *0, 0.28, 0.52, 0.76* and *1*. Insert: drying rate curves for layers composed with rigid ($\phi =1$) and soft spheres ($\phi=0$). The arrow shows the time of cracks formation for the layer of pure rigid spheres. The process of colloidal thin films drying can be subdivided into three regimes: the first one, denoted by I, is known as the constant rate period; in region II, the drying rate starts to decrease; finally, in the falling rate period denoted by III, the liquid phase breaks up into separate fractions.

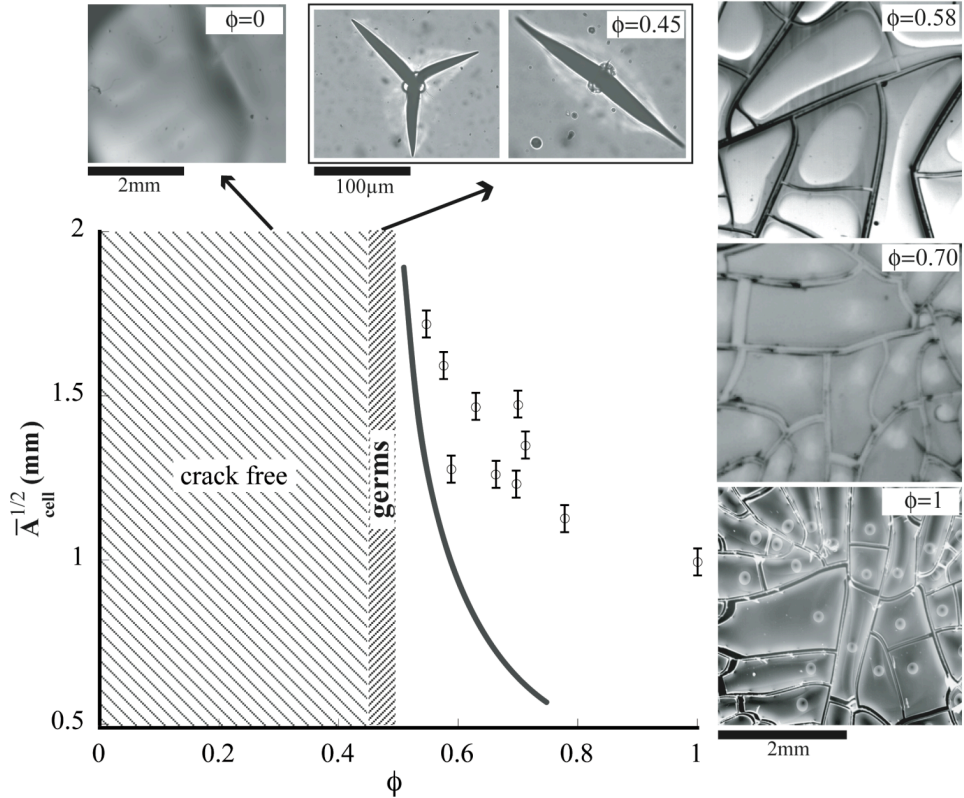

**Figure 3.** Cracks patterns quantification vs. film composition. The square root of the mean cell area, $\overline{A}_{cell}^{1/2}$ is plotted as a function of the proportion in hard spheres $\phi$, for a final film thickness $H=170\pm5\mu m$. Below a critical value of hard spheres $\phi \sim 0.45$, no crack formation could be observed. Around $\phi \sim 0.45$, isolated cracks form in the drying film. At higher volume fraction, crack formation takes place and divides the structure into polygonal adjacent cells. Images of cracks patterns for $\phi=0.58, 0.70, 1$ are shown at the final stage of the drying process: bright regions inside each polygonal cell correspond to the adhering region at the final stage of the delamination process. Drying process was investigated at $RH=37\%$. The dashed line corresponds to the theoretical adjustment.

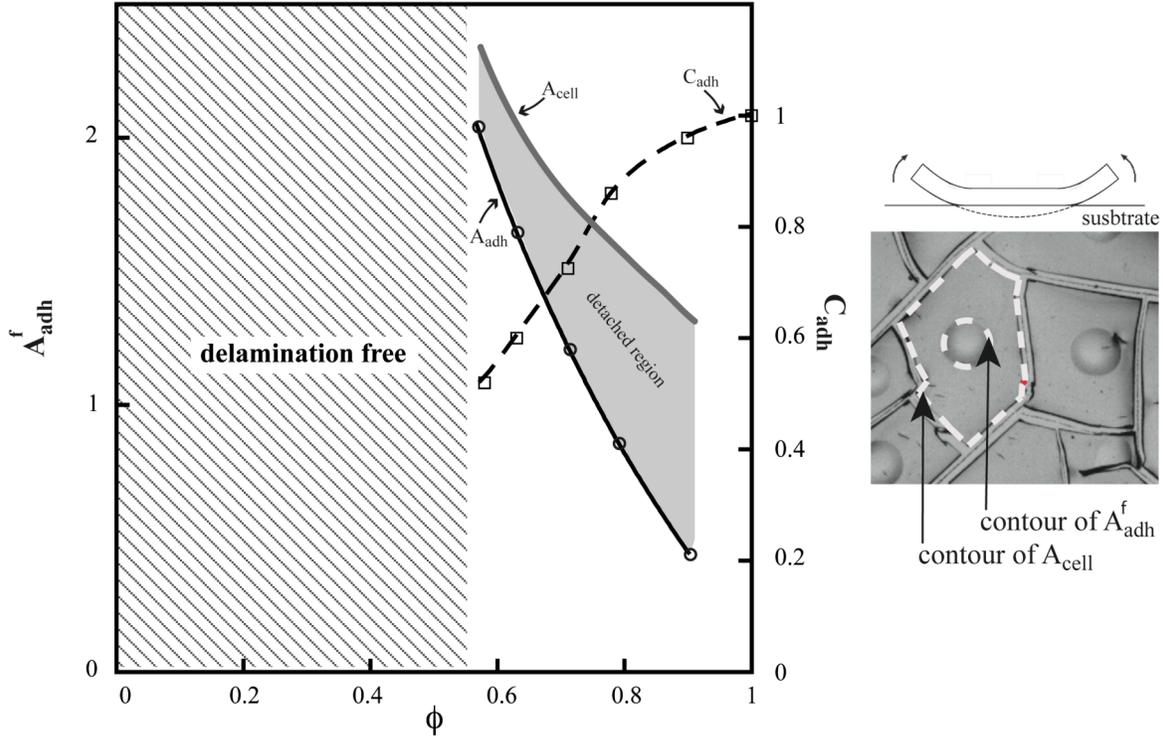

**Figure 4.** Comparison of the buckle morphologies for different proportions in hard spheres, $\phi$, at the final stage of the delamination process. (○) Measurements of the final *adhering* surface area, $A^f_{adh}$, vs. the proportion in hard spheres, $\phi$; The grey region expanding above the adhering surface area corresponds to the detached part of the film. (□) Measurements of the final circularity, $C_{adh}$, vs. the proportion in hard spheres, $\phi$ ($C_{adh}$ is defined as $4\pi \, (area)/(perimeter)^2$, area and perimeter being measured on a given surface). Lines are guides for the eyes.

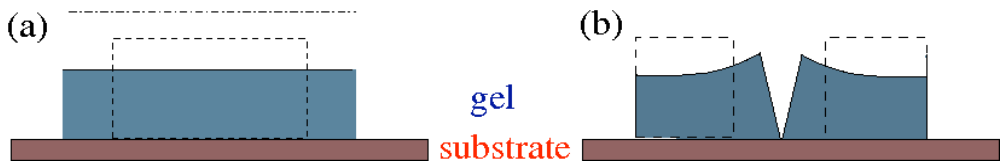

**Figure 5.** Schematic sectional view of the gel layer and substrate. (a) The initial sample (up to the dot-dashed horizontal line) is dried down to the thickness shown by a filled thick rectangle: the material cannot change its dimensions along the horizontal directions because the gel is attached to the substrate. If the gel were not attached to the substrate, it would take the form of the dashed open rectangle. (b) Cutting the material into two blocks : a crack opens by creating the V-shaped new surface when the gel is attached to the substrate. If the gel were not attached to the substrate, it would take the form of the dashed open rectangle .

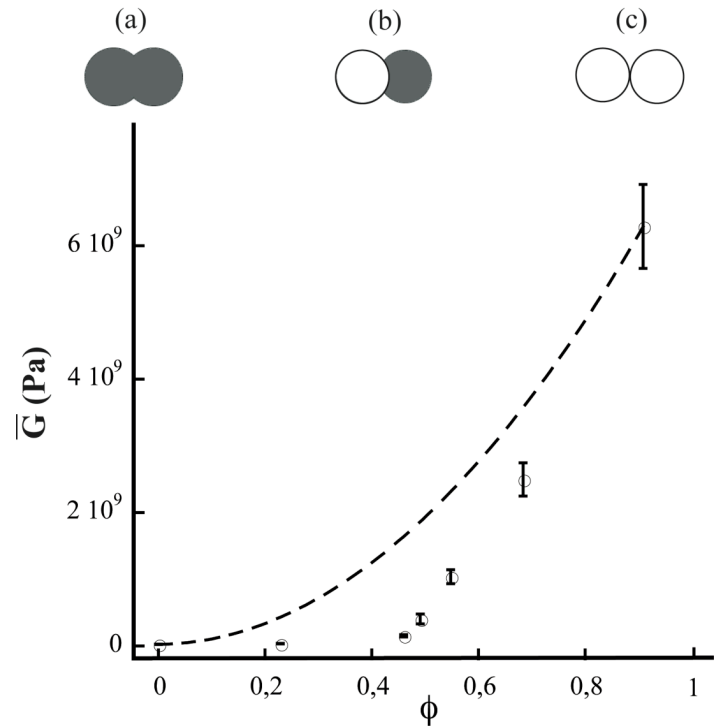

**Figure 6.** Macroscopic elastic response of the solid layer as a function of the proportion in hard spheres. Dots represent the measurements performed with micro-indentation tests, while the dashed line corresponds to the model, adapted from Russel *et al.* (2008) to the case of mixtures of hard and soft spheres.

| $\phi$ | $t_{crack}$(s) (±300s) | $\phi_{w,crack}$ (±0.03) |
|---|---|---|
| 0.52 | 11373 | 0.30 |
| 0.76 | 10863 | 0.33 |
| 1 | 10353 | 0.40 |

**Table 1:** Time $t_{crack}$ at which cracking occurs for gel of various compositions $\phi$. An estimate of the water volume fraction $\phi_{w,crack}$ when cracking occurs is also calculated. The solid volume fraction 1- $\phi_{w,crack}$ can then easily be deduced from the data.